\documentclass[10pt,twocolumn,letterpaper]{article}

\usepackage{newtxtext}
\usepackage{newtxmath}

\usepackage[
  letterpaper,
  left=0.75in,
  right=0.75in,
  top=1in,
  bottom=1in
]{geometry}

\usepackage{microtype}
\usepackage{graphicx}

\usepackage{graphicx}

\usepackage{array}

\usepackage{pifont}%
\usepackage{graphicx}
\usepackage{xcolor}
\usepackage{listings}
\usepackage{xspace}

\usepackage{xurl}

\definecolor{linkblue}{RGB}{0,70,140}

\usepackage[
]{hyperref}

\hypersetup{
  colorlinks=true,
  allcolors={blue!60!black}
}

\usepackage[nameinlink]{cleveref}

\usepackage{tabularx}
\usepackage{array}
\usepackage{diagbox}
\usepackage{makecell}

\usepackage{wrapfig}

\usepackage{booktabs}
\usepackage{multirow}
\usepackage{graphicx}
\usepackage{siunitx}
\usepackage{pifont}%
\usepackage{enumitem}

\newcommand{\cmark}{\ding{51}}%
\newcommand{\xmark}{\ding{55}}%

\newcommand{\todo}[1]{\textcolor{red}{#1}}

\newcommand{\codename}{{\sc Devlore}\xspace}

\newcommand{\loc}{LoC\xspace}
\newcommand{\sm}{monitor\xspace}
\newcommand{\Sm}{Monitor\xspace}

\newcommand{\dtuple}{\texttt{(devID,function)}\xspace}

\newcommand{\shweta}[1]{\textcolor{purple}{}}
\newcommand{\srdjan}[1]{\textcolor{cyan}{}}
\newcommand{\supraja}[1]{\textcolor{blue}{}}
\newcommand{\fried}[1]{\textcolor{green}{}}

\newcommand{\arxiv}[1]{}

\newcommand{\gptc}{GPT$_{c}$\xspace}
\newcommand{\gptd}{GPT$_{d}$\xspace}

\newcommand{\stables}{S2 tables\xspace}
\newcommand{\stable}{S2 table\xspace}

\newcommand{\rsiattachdev}{\texttt{rsi\_attach\_dev}\xspace}
\newcommand{\rsidetachdev}{\texttt{rsi\_detach\_dev}\xspace}
\newcommand{\rmidevfinalise}{\texttt{rmi\_dev\_finalize}\xspace}

\newcommand{\smcprotint}{\texttt{smc\_prot\_int}\xspace}

\newcommand{\smcackint }{\texttt{smc\_ack\_int}\xspace}

\newcommand{\smcgicconfig}{\texttt{smc\_gic\_config}\xspace}

\newcommand{\rmigranuledelegate}{\texttt{rmi\_granule\_delegate}\xspace}
\newcommand{\rmigranuleundelegate}{\texttt{rmi\_granule\_undelegate}\xspace}
\newcommand{\rmidatacreate}{\texttt{rmi\_data\_create}\xspace}
\newcommand{\rmidatadestroy}{\texttt{rmi\_data\_destroy}\xspace}

\newcommand{\config}{configuration\xspace}
\newcommand{\configs}{configurations\xspace}
\newcommand{\cvm}{CVM\xspace}
\newcommand{\cvms}{CVMs\xspace}

\crefname{figure}{Fig}{Fig}
\crefname{section}{Sec}{Sec}
\crefname{appendix}{Appx}{Appx}
\crefname{table}{Tab}{Tab}
\crefname{listing}{Lst}{Lst}
\usepackage{graphicx}
\usepackage{svg}

\SetEnumitemKey{myenum}{leftmargin = *}

\newlist{genum}{enumerate}{1}
\setlist[genum]{label*=G\arabic*:~,ref=G\arabic*}
\makeatletter
\newcommand\myitem[1][]{%
  \if\relax\detokenize{#1}\relax
    \item\relax
  \else
    \protected@edef\@currentlabel{\textcolor{black}{G$_{#1}$}} %
    \item[\textbf{G$_{#1}$:}]
  \fi}
\makeatother

\newcommand\newparagraph[1]{%
  \noindent\textbf{#1}%
}

\newcommand{\B}{\textbf{B}\xspace}
\newcommand{\I}{\textbf{I}\xspace}

\newcommand{\us}{$\mu\text{s}$}

\usepackage{tikz}

\makeatletter
\newcommand{\circnum}[1]{%
\hyperref[fig:gpu:latency]{%
  \tikz[baseline=(char.base)]{%
    \node[draw,circle,inner sep=0pt,minimum size=2.5ex] (char) {%
      \vphantom{A}\@roman{#1}%
    };%
    }%
  }%
}
\makeatother

\newcommand{\step}[1]{\hyperref[fig:design:overview]{Step #1}}
\newcommand{\steps}[1]{\hyperref[fig:design:overview]{Steps #1}}
\newcommand{\stepsNoText}[1]{\hyperref[fig:design:overview]{#1}}
\usepackage{caption}
\makeatletter
\newcommand{\repeatthanks}[1]{%
  \textsuperscript{\@fnsymbol{#1}}%
}
\makeatother

\begin{document}
\raggedbottom
\title{\codename: Device Interrupt Protection for Confidential VMs}

\author{
  \rm Andrin Bertschi\thanks{These authors contributed equally to this work.}
 \; Supraja Sridhara\repeatthanks{1}
 \; Mark Kuhne \\
 \; Benedict Schlüter 
 \; Friederike Groschupp 
 \; Clément Thorens 
 \; Nicolas Dutly  \\
 \; Srdjan Capkun
 \; Shweta Shinde
   \\[0.4em]
{ETH Zurich}
 }
 \date{}
 
\maketitle              %

\begingroup
\renewcommand{\thefootnote}{}
\footnotetext{This is the extended version of the paper accepted at RAID 2026.}
\endgroup

\begin{abstract}
Modern confidential computing executes sensitive
computation in an abstraction called confidential VMs and protects from
the hypervisor, host OS, and other co-resident VMs.
It has been shown that an attacker can inject 
malicious interrupts to break the confidentiality and
integrity of confidential VMs.
We present \codename, a device interrupt isolation mechanism that protects confidential VMs from interrupt manipulation attacks.  
Our design employs a delegate-but-check strategy by offloading interrupt 
management to the hypervisor, but adds correctness checks in the
trusted software.
We prototype our design on Arm Confidential Computing Architecture (CCA). 
We evaluate it on Arm FVP to demonstrate four diverse devices
attached to confidential VMs and report costs on a Rock 5B board.
Our case studies show the feasibility of real-world use cases and that \codename incurs minimal overheads of 0.06\% for typical integrated GPU applications.

\end{abstract}

\section{Introduction}

Arm-based platforms are deployed in consumer electronic devices (e.g., phones, cars, laptops), edge devices, and the cloud.
Applications running on these platforms typically use integrated devices (e.g., display, sensors, integrated GPUs) to perform  sensitive tasks (e.g., inference, authentication) on-device or at edge.
Cloud providers offer integrated accelerators connected to Arm CPUs (e.g., Grace-Hopper and Grace-Blackwell integrate Arm processors with state-of-the-art GPUs~\cite{grace-blackwell-azure,nvidia-digits}).
In these settings, it is important to safeguard users' sensitive computation on the platform from untrusted software (e.g., hypervisor executing on personal devices) and untrusted providers (e.g., cloud management).

Confidential computing enables users to execute their computation in isolation from privileged untrusted software.
Arm Confidential Computing Architecture (CCA), similar to Intel TDX and AMD SEV-SNP, provides a confidential virtual machine (\cvm) abstraction with a new hardware extension called realm management extensions (RME)~\cite{tdx,sev-snp,cca}.
CCA isolates \cvms from each other and from untrusted hypervisors and host OSes.
To achieve this, CCA introduces two new worlds called the realm and the root world. 
The root world has the highest privilege and exclusively executes the trusted firmware including the \sm.
The realm world houses the \cvms and a Realm Management Monitor (RMM).
These new worlds are in addition to two existing worlds: the normal/non-secure world for the host hypervisor and the secure world for trusted OS and apps.

Arm CCA-based \cvms executing in the realm world may need access to integrated devices on both consumer devices and the cloud.
For example, to use a keypad or biometric sensor to perform  authentication or to use an integrated GPU for computation on sensitive data in the cloud.
However, CCA prohibits devices from accessing \cvm memory.
Recent works address this gap by allowing devices and \cvms to directly access each others' memory. They enforce memory isolation guarantees to ensure that device access does not compromise the \cvm security~\cite{cage,acai,portal}. 
On the other hand, recent works have shown that interrupt attacks from a malicious hypervisor can compromise the \cvm execution~\cite{heckler-usenix,wesee-oakland}.

To address this gap, we aim to enforce interrupt isolation---malicious interrupts injected by the attacker are never delivered to the CVMs, while ensuring that benign device interrupts are delivered correctly. 
Interrupt isolation presents challenges that are different from device memory isolation. 
For memory isolation, each device belongs to exactly one \cvm. 
So, once the device is attached via memory setup, the hardware enforces the isolation without further software intervention. 
In contrast, the interrupt controller and its state is one resource that needs to be shared and multiplexed between the hypervisor and all the \cvms. 
Further, we not only need to isolate the interrupt \config but also runtime delivery and acknowledgment.

We consider several potential designs to achieve this 
primitive of interrupt isolation. Our design exploration shows that obvious solutions are insufficient. Further, since enforcing the primitive entails reasoning about the entire interrupt lifecycle (registration, physical to virtual conversion, delivery, priority, acknowledgment), we have to weigh the design options for each phase.
We isolate both physical and virtual interrupts without any software changes to existing device drivers. Our insight is to delegate the interrupt management and delivery to the untrusted hypervisor while the trusted software checks if the hypervisor misbehaves.
Our approach minimizes software changes because of this delegate-but-check strategy while maintaining low performance overheads.

We prototype our design, called \codename, on the Fixed Virtual Platform (FVP),
a CCA-enabled emulator provided by Arm to demonstrate feasibility, correctness, and compatibility with hardware specification and software stack \cite{arm-fvp}.
Since CCA is not yet publicly available in CPUs~\cite{cobalt-azure,cobalt-arm,opencca}, we retrofit \codename changes to a Rock 5B board based on OpenCCA~\cite{opencca}.
\codename changes to the hypervisor, firmware, and the kernel (host and guest) are minimal ($\approx$7.5k\,LoC), and without device driver changes.

We test the functional correctness of \codename using our prototype on the FVP with four devices of varying complexity and isolation requirements: GPU, UART, Keyboard/Mouse, and LEDs/Buttons.
Further, our three case studies on the FVP show that \codename has non-invasive impact on applications (IRC, terminal browser, text editor).
Our measurements on the Rock 5B board performance prototype show that \codename overheads are minimal. 
First, we run interrupt stress workloads using Mali GPU and UART on the Rock 5B board. 
Our experiments show that under sustained interrupt load, \codename only incurs overheads up to 1\%.
Next, our case study using the glmark2 GPU benchmark that models typical integrated GPU applications shows that \codename overhead drops to only 0.06\%. 
\codename is available at~\cite{devlore-sourcecode}.

\vspace{0.5em}

\newparagraph{Contributions.}
The main contributions of this paper are:
\begin{itemize}[itemsep=5pt]
\item{\bf \em Interrupt Isolation.} 
    We design interrupt isolation to bridge the gap between memory and interrupts for integrated devices in CVMs on Arm CCA.
    
    \item{\bf \em Delegate-but-check.} 
    Our design, \codename, delegates the interrupt management to the untrusted hypervisor while the trusted software checks its correctness.

\item{\bf \em Devices.} \codename isolates device interrupts with no modifications to applications or device drivers and limited overheads. We demonstrate \codename with devices and case studies evaluated on both the Arm FVP and an Arm board.
\end{itemize}

\section{Arm CCA and Interrupt Management}
\label{sec:background}

We introduce Arm CCA, explain the interrupt behavior of integrated devices, and how they are virtualized for \cvms.
\begin{table}[t]
  \centering
      \caption{Arm CCA memory access rules. The columns (From) denote the security state of the processing element that originates the memory access request. The rows (To) denote physical address spaces. }
  \footnotesize
  \begin{tabular*}{\columnwidth}{@{\extracolsep{\fill}}lllll@{}}
    \toprule
 \textit{To/From}
         & \textbf{Root} & \textbf{Realm} & \textbf{Normal} & \textbf{Secure} \\
         \midrule
        \textbf{Root} & Yes & No & No & No\\
        \textbf{Realm} & Yes & Yes & No & No \\
        \textbf{Normal} & Yes & Yes & Yes & Yes \\
        \textbf{Secure} & Yes & No & No & Yes\\
        \bottomrule
    \end{tabular*}
        \label{appx:tab:cca-pas-access}
\end{table}

\begin{figure}
  \centering
\includegraphics[width=1\linewidth, trim=0cm 0cm 0cm 0cm]{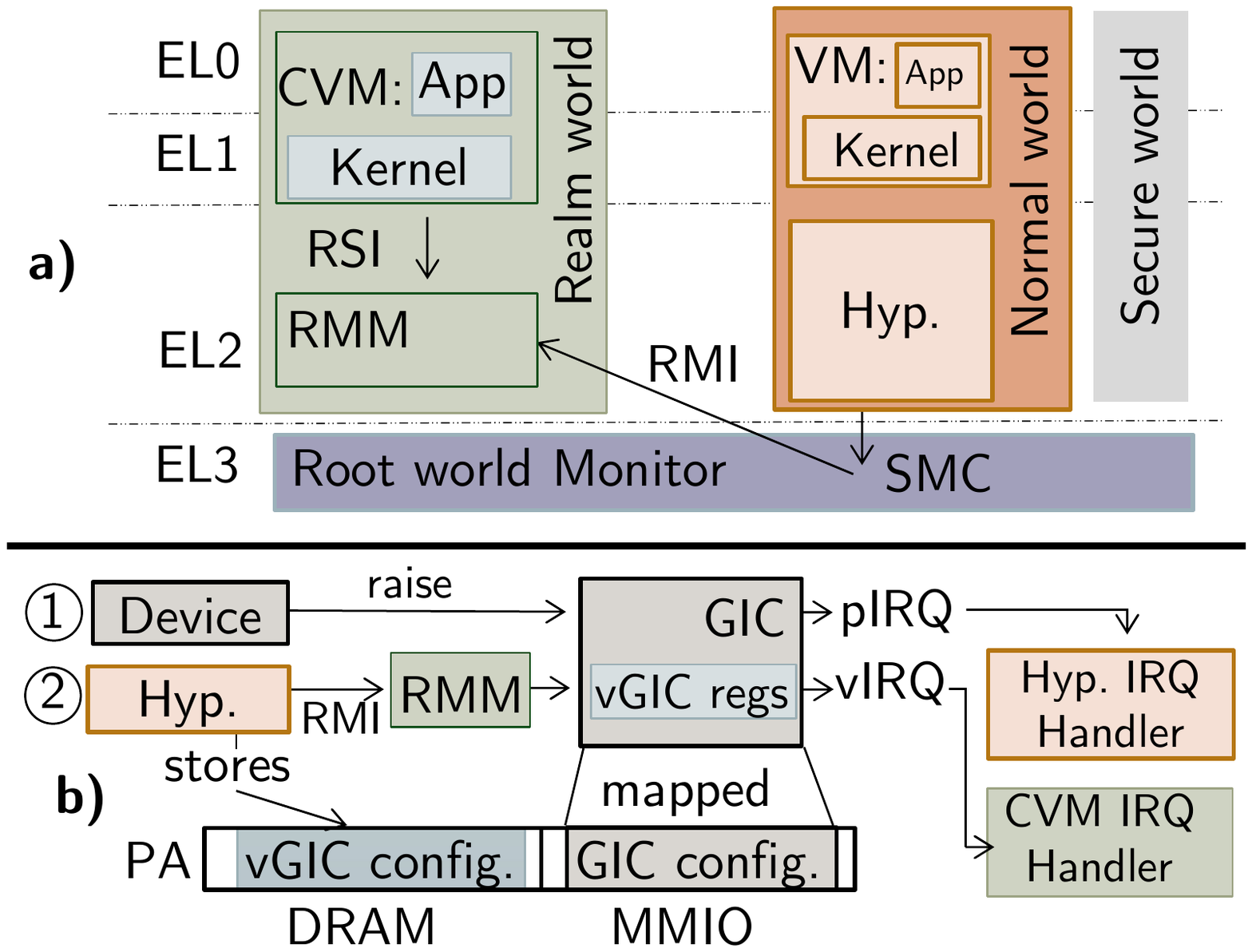}
    \caption{
    (a) Arm CCA. (b) Arm interrupt architecture with CCA. 1: Device raises physical interrupt (pIRQ) to hypervisor. 2: Hypervisor invokes RMM with vGIC configuration to virtualize CVM interrupts.
       }
    \label{fig:background}
\end{figure}

\label{ssec:arm-cca}
\newparagraph{Arm CCA.} To enable Arm CCA, the Arm ISA adds support for Realm Management Extensions (RME).
RME enables computation to execute in one of four worlds: Normal, Realm, Secure, and Root world.  
These worlds are isolated from one another and have distinct physical address spaces (\cref{appx:tab:cca-pas-access}).
RME enforces this
isolation through hardware mechanisms called Granule Protection Checks (GPC).
GPCs look up Granule Protection Tables (GPTs), which map physical addresses to the worlds. GPTs are stored in protected root world memory, and programmed by the root world Monitor. 
Apart from the worlds, the Arm architecture enables privilege levels called exception levels (EL) from EL0 to EL3.
Typically, EL0 executes apps, EL1 operating system kernels, and EL2 hypervisor.
The root world executes the Monitor at EL3, the highest privilege level.
Software in EL2 can trigger a world switch by issuing a Secure Monitor Call (SMC) to the \sm
(\Cref{fig:background}.a).

\newparagraph{CVMs.} Confidential VMs (\cvms) execute in realm world EL0-1 under the oversight of the Realm Management Monitor (RMM) that executes in realm world EL2.
The RMM does not actively manage \cvm lifecycle.
Instead, it validates and enforces operations requested by the normal world hypervisor.
For this, it exposes the Realm Management Interface (RMI) to the hypervisor via SMC calls.
The hypervisor then uses RMIs to handle CVM lifecycle (creation, expansion, deletion), and for scheduling.
The RMM isolates the CVMs using stage-two MMU translation tables (S2 tables).
Similar to the RMIs, the RMM exposes the Realm Service Interface (RSI) to its CVMs to handle their requests.

\newparagraph{Arm Interrupt Management.}
The Generic Interrupt Controller (GIC) delivers physical interrupts to CPU cores based on its \config.
The hypervisor accesses the GIC through a memory-mapped configuration  interface and can
selectively enable/disable interrupts, set interrupt affinities, and assign interrupt priorities. 
With virtualization, guest VMs can only handle virtual interrupts.
So, the hypervisor virtualizes physical interrupts from the GIC  by programming the virtual GIC (vGIC).
The vGIC provides each VM with a virtual view of the interrupt controller. 
To inject a virtual interrupt, the hypervisor programs up to $n$ vGIC system registers, where $n$ depends on the GIC implementation. 
If multiple interrupts are pending, the hypervisor programs the vGIC with the highest priority interrupts first. 
On VM entry, the vGIC delivers the programmed virtual interrupts to the guest.

\newparagraph{CVM Interrupt Handling.} %
Under standard CCA, the untrusted normal world hypervisor remains responsible for managing and virtualizing
\cvm interrupts.
When a physical device interrupt arrives while a \cvm is running, the CPU must not switch directly to the hypervisor, as this could leak \cvm state.
Instead, the interrupt first traps to the RMM, which saves the \cvm state and only then forwards control to the hypervisor (see (1) in \cref{fig:background}.b).
For security, CCA prevents the hypervisor from directly programming the \cvm vGIC system registers.
To inject virtual interrupts into the \cvm, the hypervisor must invoke the RMM with a vGIC configuration, and the RMM programs the vGIC on its behalf, before entering a \cvm (see (2) in \cref{fig:background}.b).
In its current design, the RMM cannot verify if these virtual interrupts correspond to genuine physical interrupts.
As a result, the hypervisor can inject arbitrary interrupts into a \cvm.

\section{Problem and Insights}
\label{ssec:interrupt-attacks}
We motivate the need for interrupt~isolation.

\subsection{Injecting Malicious Device Interrupts}

We discuss how an attacker can inject fake interrupts to mount interrupt attacks.

\newparagraph{Recent Interrupt-Based Attacks.}
Recent works have shown that a malicious hypervisor can use interrupt injections to compromise \cvm security~\cite{heckler-usenix,wesee-oakland}. 
They observe that confidential computing solutions (e.g., TDX, SNP, CCA) delegate \cvm management (e.g., scheduling, memory management, interrupt management) to the untrusted hypervisor. 
To manage interrupts and virtualize them for \cvms, the hypervisor controls the  interrupt controller that delivers interrupts to the cores. 
So, the hypervisor can program the GIC to inject interrupts directly into the \cvms at any point during their execution. %
Since the \cvm is not aware of this, similar to Iago attacks~\cite{checkoway2013iago}, it breaks execution integrity.
Next, we discuss potential threats to \cvms connected to devices. 
\begin{figure}
        \centering
     \includegraphics[width=0.85\columnwidth, trim=0cm 3.5cm 0cm -0.5cm
     ]{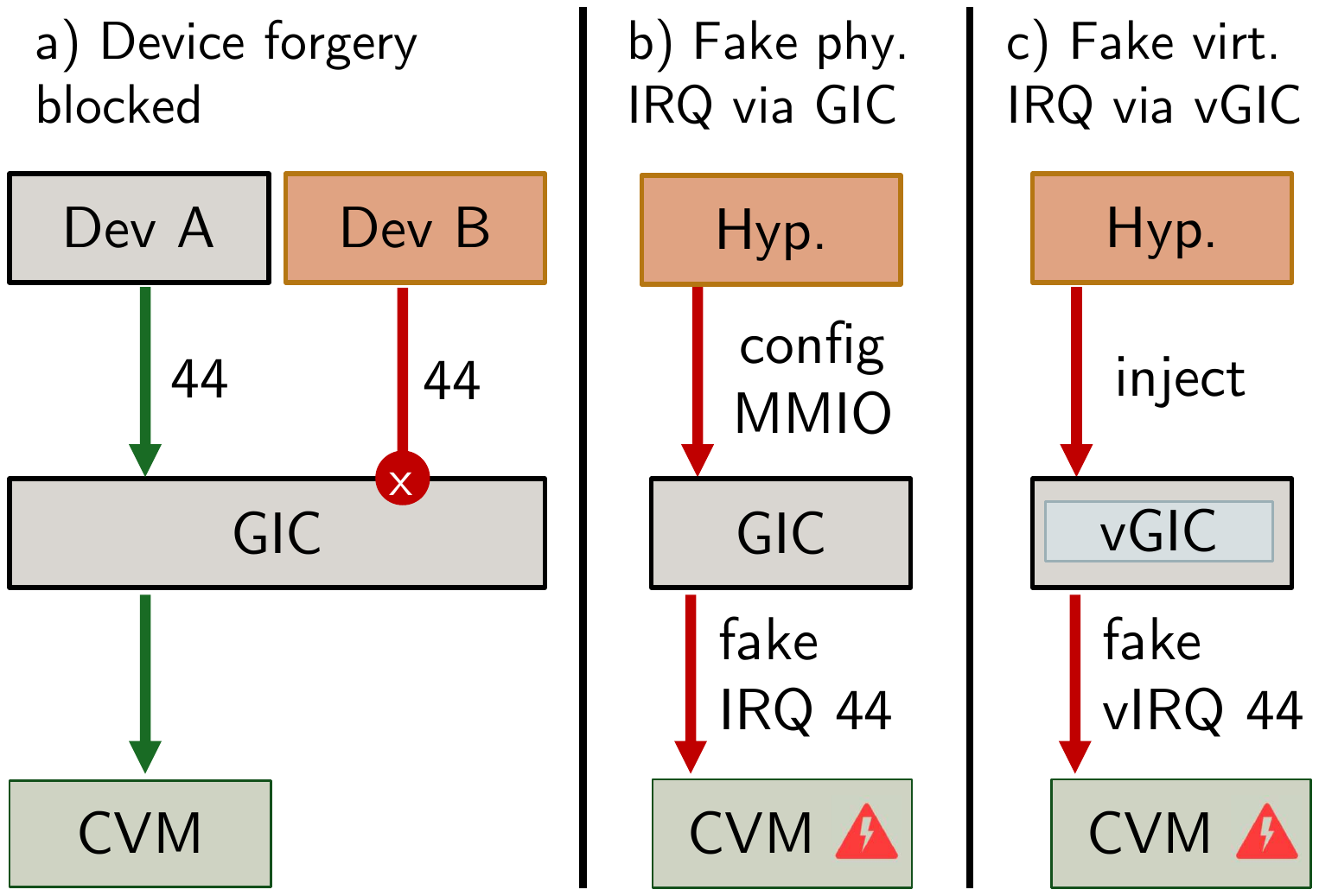}
         \caption{Attacks. (a) Devices cannot change their physical interrupt numbers.
         Hypervisor injects fake physical (b) and virtual (c) interrupt (IRQ 44). 
         }
    \label{fig:attacks}
\end{figure}

\newparagraph{Untrusted Devices.}
On any given Arm platform, an integrated device can only assert its designated interrupt line. 
So, integrated devices cannot send physical interrupts to fake interrupts from other devices that might be attached to \cvms (see \cref{fig:attacks}.a).
Therefore, even if an attacker controls an integrated device they cannot use it to inject fake interrupts. 
Next, we discuss threats from untrusted management software (e.g., hypervisors) via interrupts~\cite{heckler-usenix,wesee-oakland}. 

\newparagraph{Untrusted Management Software.}
The hypervisor manages the interrupts for the \cvms and can abuse it in two ways: 
(i) configure the GIC to trigger fake  physical interrupts (\cref{fig:attacks}.b); and
(ii) use virtual interrupt injection through the vGIC system registers to deliver fake virtual interrupts of either type (\cref{fig:attacks}.c). 
Next, we analyze device drivers to understand how they handle interrupts. 

\subsection{Driver Examples}
\label{ssec:driver-analysis}

To understand how device drivers handle interrupts, we manually analyzed the open-source drivers of popular integrated devices in Linux.

\newparagraph{Counter.}
The Linux kernel counter subsystem implements an interrupt-driven counter. 
This counter can use any interrupt source (e.g., temperature sensor, proximity sensor) as an event source and increments a counter when it receives an interrupt (\cref{lst:ctr}).
The counter implementation can be used by user-space applications to be notified when the counter is incremented, e.g., when an event has occurred for a fixed number of times (Line 5). 
Assume that the interrupt source for this counter is connected to a \cvm and this counter is used by a \cvm application.
The untrusted hypervisor can maliciously inject interrupts, trigger this interrupt handler, and increment the counter to trigger the \cvm application event condition eventually. 
Thus, using malicious interrupt injection, the hypervisor successfully tricks the \cvm into believing an event occurred when, in reality, it did not, breaking its execution integrity.

\begin{lstlisting}[language=C, caption={Linux interrupt-driven counter.}, label=lst:ctr,captionpos=b]
irqreturn_t interrupt_cnt_isr(
   int irq, void *dev_id)
{
   struct counter_device *ctr = dev_id;
   struct interrupt_cnt_priv *priv 
      = counter_priv(ctr);
   atomic_inc(&priv->count);
   counter_push_event(ctr,
      COUNTER_EVENT_CHANGE_OF_STATE, 0);
   return IRQ_HANDLED;
}
\end{lstlisting}

\newparagraph{Wireless Device Boot-up.}
Qualcomm WCNSS chip is used in a wide range of devices with Arm cores for wireless communications and is vulnerable to the untrusted hypervisor interrupt injection attacks. 
Concretely, the WCNSS driver used for verified firmware loading expects interrupts to indicate when the device  firmware is successfully loaded and ready to use (see \cref{lst:wcnss})~\cite{linux-wcnss-source}. 
  \begin{lstlisting}[language=C, caption={Ready interrupt handler in WCNSS driver.}, label=lst:wcnss,captionpos=b]
irqreturn_t wcnss_ready_interrupt(
   int irq, void *dev)
{
   struct qcom_wcnss *wcnss = dev;
   /* Change to ready: */
   complete(&wcnss->start_done); 
   return IRQ_HANDLED;
}
\end{lstlisting}
If the hypervisor injects an interrupt to trigger this handler before the device is ready, it tricks the driver into transitioning the device state to ready. 
This may have undesirable effects (e.g., \cvm may use an outdated or corrupted WiFi firmware image).
Thus, the hypervisor may break the driver execution integrity using only malicious interrupt injection.

\subsection{Threat Model and Scope}
We trust the platform and all of its components, including the interrupt controller (GIC). 
Like previous works~\cite{portal,cage}, we assume that the hardware integration of the devices on the platform conforms to the specification and is bug-free, i.e., devices are only accessible through the specified address ranges and can only assert their assigned interrupts. 
We assume that all software executing in the normal and secure worlds is untrusted. 
We trust the CCA hardware, assume that all trusted CCA software (i.e., the RMM and \sm) is implemented according to CCA specifications, and assume mutual distrust between \cvms. 
For a given \cvm, we consider all software executing in \cvm (i.e., operating system, device drivers and runtimes, applications) as trusted. 
Protecting against side-channels and microarchitectural attacks is orthogonal and we consider them out of scope.

\newparagraph{Problem Statement.}
We aim to provide guarantees for interrupts from integrated devices delivered to a \cvm. 
Concretely, we ensure that the device interrupts delivered to a \cvm in the presence of an untrusted hypervisor is equivalent to interrupts that would be delivered if the hypervisor was trusted. 
This guarantees that fake interrupts injected by a malicious hypervisor cannot compromise \cvm security.

\section{Potential Solution}
\label{ssec:potential-mitigations}

Our analysis of the device drivers in the Linux kernel shows that in addition to pure interrupts, the devices use two other types of event notification mechanisms: (1) only poll memory-mapped regions, or (2) use interrupts in conjunction with memory-mapped region reads.
In both these cases, the interrupt handler reads or checks isolated MMIO memory before processing the interrupt. 
This isolated MMIO memory is not accessible to the malicious hypervisor. 
Therefore, if the drivers check some state in the device  MMIO memory before processing the interrupts, they can prevent the interrupt attacks from~\cref{ssec:interrupt-attacks}.

If all device drivers adopt one of the two mechanisms above and the MMIO regions are isolated, they can thwart malicious interrupt injection. 
However, this solution is infeasible in some cases, such as the interrupt-driven counter, which only depends on an interrupt source and is device agnostic. 
Because of the device-agnostic nature of this counter, reading from an MMIO space is not possible, as different devices have different MMIO spaces and behaviors. 
Therefore, supporting this interrupt-driven counter in a \cvm such that the hypervisor cannot corrupt it requires architectural guarantees for interrupt isolation. 
For devices where such memory-mapped reads would be possible (e.g., Qualcomm WCNSS), doing so may incur hardware changes to the device to write to the register. 

To adopt this solution, we will need to change any device that simply raises an interrupt when an event occurs to also write to a memory-mapped region.
Further, we will need to change all existing device drivers where such a check is feasible. 
Doing this will incur significant changes to the hardware and manual driver patching. 
More importantly, \emph{even one device left unchanged or one driver left unpatched} will leave the \cvm vulnerable to interrupt attacks.
We observe that such invasive hardware and driver changes are not necessary if we can provide architectural guarantees for interrupt delivery. 
Therefore, in \codename we take a principled approach to ensure isolated interrupt delivery to the \cvms.

\section{Solution Overview}
\label{sec:overview}
\begin{figure*}[t]
    \centering
        \includegraphics[width=0.98\textwidth, trim=0cm 7cm 2cm 6cm]{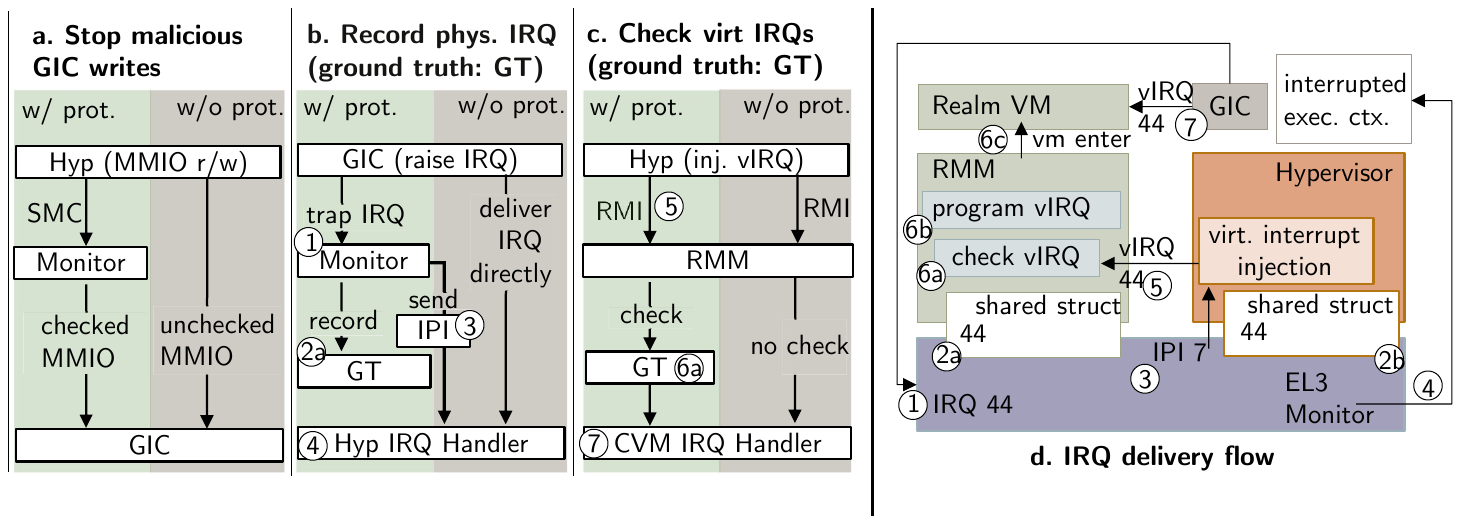}
    \caption{\codename solution overview.
        (a-c): \codename ensures interrupt isolation by preventing malicious GIC writes,
        recording authentic physical interrupts as ground truth (GT), and checking hypervisor-injected virtual
        interrupts against GT.
        (d): Interrupt delivery flow.
        Steps (1,2ab,3,4) record the physical interrupt,
        Step (5) virtualizes it,
        and Steps (6a,b,c,7) check and deliver it.
        Example for IRQ 44: (1) IRQ 44 traps to the \sm;
        (2a,b) the \sm writes 44 to the RMM and hypervisor shared data structures;
        (3) it notifies the hypervisor via software interrupt 7 (IPI);
        (4) it returns to the interrupted application;
        (5) the hypervisor schedules the guest and sends the virtual interrupt configuration to the RMM;
        (6a,b) the RMM checks and programs vGIC system registers;
        (6c) it enters the CVM; 
        (7) the GIC fires virtual interrupt 44.
    }
    \label{fig:design:overview}
\end{figure*}

First, we provide an intuition behind \codename and outline its three steps. The three steps are visualized in \Cref{fig:design:overview}a-c.
Recall that on Arm platforms, untrusted integrated devices cannot fake physical interrupts from other devices as they can only assert their designated interrupt lines.
The hypervisor can inject physical interrupts only by writing to the GIC directly. 
So, by constraining the hypervisor capability to maliciously write to the GIC, we guarantee that all registered physical device interrupts originate solely from their respective devices. 
Using this observation, in \codename we first \textbf{prevent malicious GIC writes (\cref{sec:overview:prevent:writes})} from the hypervisor to ensure physical device interrupt authenticity.
However, physical interrupt authenticity alone is not sufficient. 
We also need to isolate the virtual interrupts that are delivered to the \cvm. 
In CCA, the hypervisor performs interrupt management. 
So, the hypervisor can always inject fake virtual interrupts to the \cvms.
To prevent malicious virtual interrupt injection, we only allow registered virtual interrupt injection if there was a corresponding authentic physical interrupt from the device.
    \cvms must \emph{register} specific device interrupts to isolate from the hypervisor using \codename. 
We call these interrupts the \cvm \emph{registered interrupts}. 
To isolate these interrupts, \codename  \textbf{records all registered physical device interrupts (\cref{ssec:record-phy-intr})}  to create a ground truth.
Then, when the hypervisor injects virtual interrupts, \codename \textbf{checks the virtual interrupts (\cref{ssec:checking-virtual-interrupts})}
against this ground truth and ensures that every registered interrupt delivered to the \cvm corresponds to an authentic physical interrupt.

\subsection{Prevent Malicious GIC Writes}
\label{sec:overview:prevent:writes}
Our first step is to prevent a malicious hypervisor from injecting  physical interrupts by writing directly to the GIC (see \Cref{fig:design:overview}.a).
\codename stops the hypervisor from directly writing to the GIC memory by marking the GIC memory as root world and using the \sm to check any updates to the GIC \config. 
This ensures the physical interrupt authenticity, i.e., any physical device interrupt that arrives at the cores is guaranteed to only  be from the device.

\subsection{Recording Physical Interrupts}
\label{ssec:record-phy-intr}
After establishing the authenticity of physical device interrupts, we need a way to record these when they arrive at the cores
(see \Cref{fig:design:overview}.b).
By default, the GIC sends a physical interrupt to any core that is active. 
In CCA, all these interrupts are forwarded to the hypervisor for management. 
So registered interrupts must first be delivered to trusted software (e.g. \cvm, RMM, \sm), recorded there, and only then forwarded to the hypervisor.
Below, we discuss these potential approaches that enable such recording and present \codename's approach. 

\newparagraph{Option 1: Deliver to \cvm.}
We can consider delivering the physical interrupt directly to its \cvm. 
However, this is not feasible as the \cvm runs in a virtualized environment and cannot directly handle physical interrupts. 
Instead, on receiving the physical interrupt the hardware forces the \cvm to exit and traps to the RMM (in EL2). 
Therefore, the \cvm cannot record these physical interrupts. 

\newparagraph{Option 2: Deliver to RMM.}
Physical interrupts always trap to EL2. 
We can ensure that the registered physical interrupts are always delivered to the RMM which executes in realm EL2. 
To isolate interrupt delivery to RMM, the interrupt must only target cores executing in the realm world.
However, ensuring this is not straightforward. 
In CCA, the hypervisor is responsible for all scheduling decisions. 
Thus, any core may be executing in any world at any time.
To ensure reliable physical interrupt delivery to the RMM, we need to get around this dynamic scheduling behavior. 
For this, we can pin a core that always executes the RMM and program the GIC to only deliver the registered device interrupts to this core~\cite{castes2024sharing}. 
With this the RMM can record all physical interrupts that arrive, and then forward the interrupt to the hypervisor. 
While this approach allows us to record the interrupts, 
it wastes a core (pinning) and creates a bottleneck.

\newparagraph{Option 3: Change the GIC Hardware.}
In Option 2, we require core pinning because the GIC cannot route specific interrupts exclusively to cores in the realm world.
This could be avoided by extending the GIC hardware with a realm-world mechanism analogous to the secure world.
The secure world can configure the GIC to deliver specific interrupts only to cores in the secure world.
However, the GIC is a highly optimized piece of hardware and to propose changes to it, we need access to the hardware and tools which we do not have. 
More importantly, these hardware changes will not be sufficient to ensure virtual interrupt isolation (as discussed in~\cref{ssec:checking-virtual-interrupts}).
Therefore, \codename records interrupts without hardware changes.

\newparagraph{\codename Solution: Deliver to \Sm. }
By default, the GIC provides a mechanism to ensure that specific interrupts are only delivered to EL3,
 i.e., the \sm.
We configure the GIC via the \sm so that all registered interrupts are delivered to the \sm (\step{1 in \Cref{fig:design:overview}}.b).
Upon arrival, the \sm records the physical interrupt in a realm data structure (\step{2}),
forwards it to the hypervisor (\step{3}), and returns the interrupt context (\step{4}).
This data structure serves as the ground truth for virtual interrupts.

\subsection{Checking Virtual Interrupts}
\label{ssec:checking-virtual-interrupts}
After authenticating recorded physical interrupts, we need to isolate and check the virtual
interrupts (\Cref{fig:design:overview}.c).
We discuss the main design options. 

\newparagraph{Option 1: Move Interrupt Management to the RMM.}
CCA employs a delegate-but-check strategy for all \cvm management. 
So, it lets the hypervisor perform interrupt management for \cvms. 
We can move the virtual interrupt management to the RMM such that the hypervisor has no opportunity to perform malicious interrupt delivery.
Now, the RMM is responsible for all virtual interrupt management, including the non-registered interrupts (e.g., timers, IPIs). 
Moving this to the RMM is non-trivial because these interrupts are tightly coupled with hypervisor \cvm scheduling and would require invasive changes.

Instead, we can move only the registered interrupt management to the RMM. 
To do this, the monitor records the physical interrupts and then forwards all registered interrupts to the RMM. 
Then, when the \cvm is scheduled, the RMM can program the \cvm vGIC system registers to inject the virtual interrupts. 
We observe that the same vGIC system registers are also used by the hypervisor to forward other non-registered virtual interrupts (e.g., timers, IPIs) to the \cvm. 
Because the number of vGIC system registers are limited, this approach leads to contention between the RMM and the hypervisor for the same registers. 
More importantly, the RMM now has to reason about priority of interrupts for all the interrupts (registered and non-registered) that are to be sent to the \cvm . 
This further complicates the design and requires the RMM to hold some interrupts to be sent to the \cvm in a queue introducing queue management complexity. 
We conclude that this approach, while feasible, is not practical and breaks CCA delegate-but-check strategy. 

\newparagraph{\codename Solution: Check in the RMM.}
By default, CCA does not allow the hypervisor to directly program the \cvm vGIC system registers. 
Instead, every time the hypervisor schedules a \cvm, it can request the RMM to program the \cvm vGIC to inject the virtual interrupts (\step{5 in~\cref{fig:design:overview}}.c). 
In \codename, the RMM rigorously checks that the virtual interrupt injection is benign.
To perform this check, the RMM looks up the data structure with authentic physical interrupts from the \sm (\step{6a}). 
This way, the RMM can distinguish between the case where the hypervisor 
is injecting a virtual interrupt corresponding to an authentic physical interrupt versus a malicious virtual interrupt. 
Thus, \codename ensures that the hypervisor cannot inject arbitrary virtual interrupts while preserving ordering and interrupt priorities as we explain in detail in \cref{ssec:design-check-virt-int}.
\codename design ensures that even if the interrupt management remains in the hypervisor, it cannot break \cvm security (see security analysis in \cref{sec:security_analysis}).

\section{Design}
\label{sec:design}

We detail \codename interrupt isolation mechanism.

\newparagraph{Assumptions.} 
\codename uses techniques from existing works to attach devices to a 
\cvm, ensure runtime memory isolation, and detach devices from the \cvms \cite{acai,rme-da,portal}.
For security, these prior works provide the below guarantees.

\begin{genum}[labelwidth=0.75em, wide=0pt]    
    \myitem[id]\label{g:id} {\em Identity.} Attackers cannot forge device identities and physical memory used to access a device is fixed by the platform integration.
    \myitem[a]\label{g:attach} {\em Device Attach.} Devices are attached to a \cvm in a trusted state (including device reset and attestation).
    \myitem[mem]\label{g:mem} {\em Memory Isolation.} Only the device and the \cvm it is attached to can access the device~memory.
    \myitem[d]\label{g:detach} {\em Device Detach.} Detaching devices from \cvm will not compromise \cvm security (e.g., leak residual device state). 
\end{genum}

\newparagraph{Interrupt Registration.}
\codename extends the device attach process of prior works to allow \cvms to  enable \codename interrupt isolation~\cite{portal,rme-da,acai}.  
Generally, a single device may generate multiple physical interrupts for different functions. 
For example, a Mali GPU can raise a total of three interrupts, one each for power management, on page-faults, and job completion. 
By default, the hypervisor provides the virtual interrupt information to \cvms in the form of a mapping from \dtuple to virtual interrupt number. 
With \codename, the platform manufacturer provides a trusted mapping from \dtuple to platform-specific physical interrupt number. 
We observe that the tuple \dtuple is unique on any given platform. 
So, the RMM can use this tuple to securely manage interrupt registration.
Concretely, we have to ensure that although the physical interrupt numbers vary across platforms and the hypervisor may choose different virtual interrupt numbers for \cvms, the hypervisor cannot abuse this to tamper with the virtual-to-physical interrupt bindings. 

When a \cvm wants to register an interrupt, it requests the RMM with \dtuple and its virtual interrupt number.
On receiving this request, the RMM first checks that the device is attached to the requesting \cvm before proceeding. 
If the check passes, the RMM now needs the physical interrupt that corresponds to the \dtuple in the \cvm request. 
For this, the RMM looks up the trusted platform-specific mapping from \dtuple to physical interrupt numbers. 
This look up ties the \cvm-specific virtual interrupt number to the platform-specific physical interrupt number. 
The RMM uses this physical interrupt number to invoke the \sm to register the interrupt for \codename isolation. 
Once registered, the RMM stores the virtual interrupt number to facilitate the RMM checks on the virtual interrupt as explained in~\cref{ssec:design-check-virt-int}. 
Finally, the RMM adds the registration information (i.e., \dtuple, physical interrupt number, virtual interrupt number) to the CCA attestation report.

On device detach, the RMM checks if interrupt isolation was enabled for the device. 
If so, the RMM invokes the \sm to update the GIC configuration to de-register the interrupt isolation.

\subsection{Ensuring Interrupt Isolation with \codename}
\label{ssec:interrupts-design}

For registered device interrupts, \codename guarantees interrupt isolation (\config, delivery, acknowledgment) under an untrusted hypervisor.
Recall from~\cref{sec:overview} that \codename proceeds in 3 steps: establishing authenticated physical interrupts, recording the authenticated physical interrupts as ground truth, and checking virtual interrupts against the ground truth.

\newparagraph{Authenticated Physical Interrupts.}
In Arm CCA, the hypervisor manages the GIC by default and can maliciously modify its \config to inject interrupts. 
To circumvent this threat,
\codename moves the GIC \config into the root world and protects it using the \sm checks.
To maintain hypervisor functionality, \codename introduces a GIC \config interface for the hypervisor to use in the \sm (\Cref{fig:design:overview}.a).
Concretely, any hypervisor access to the GIC \config raises an access fault.
In the hypervisors fault handler, \codename invokes the monitor to perform the operation.
On each such request, the \sm checks that the hypervisor is not trying to reconfigure the GIC for a registered interrupt.
If the check passes, the \sm allows the configuration request through. 
With this, \codename can ensure that all registered physical interrupts that the GIC injects are authentic, i.e., from their respective devices.
However, physical interrupt authenticity alone does not guarantee interrupt isolation;
\codename must also verify virtual interrupt delivery.

\newparagraph{Recording Registered Interrupts.}
Recall that once \codename establishes physical interrupt authenticity, it uses this as the ground truth to check that the hypervisor benignly virtualizes them for the \cvm. 
To create the ground truth, \codename has to record all the registered physical interrupts that the GIC fires. 
For this, it configures the GIC such that all registered device interrupts trap to the \sm. 
This lets the \sm record their arrival.
In \codename, the \sm stores this information in the RMM-accessible realm memory and notifies the hypervisor that this interrupt arrived by writing to normal world hypervisor-accessible memory. 
Now, all that remains is to ensure that \codename checks this information to guarantee benign virtual interrupt~delivery. 

\newparagraph{Delegate-but-check Virtual Interrupt Management.}
As in CCA, the hypervisor still manages interrupts for the \cvms in \codename.
So, the hypervisor should virtualize the device interrupts that the \sm trapped for the \cvms.
To enable this, \codename changes the hypervisor to first check if there is any pending interrupt from the \sm every time the hypervisor executes. 
If there are pending interrupts, the hypervisor injects the corresponding virtual interrupts into the \cvm. 
According to CCA specifications, the hypervisor sends the vGIC \config to the RMM. 
With \codename, the RMM checks this \config and then programs the vGIC to fire the virtual interrupts to the \cvm, on \cvm entry.

\subsection{Checking Virtual Interrupts}
\label{ssec:design-check-virt-int}
\begin{figure}
  \centering
\includegraphics[width=0.9\linewidth, trim=0cm 4cm 1cm 2cm]{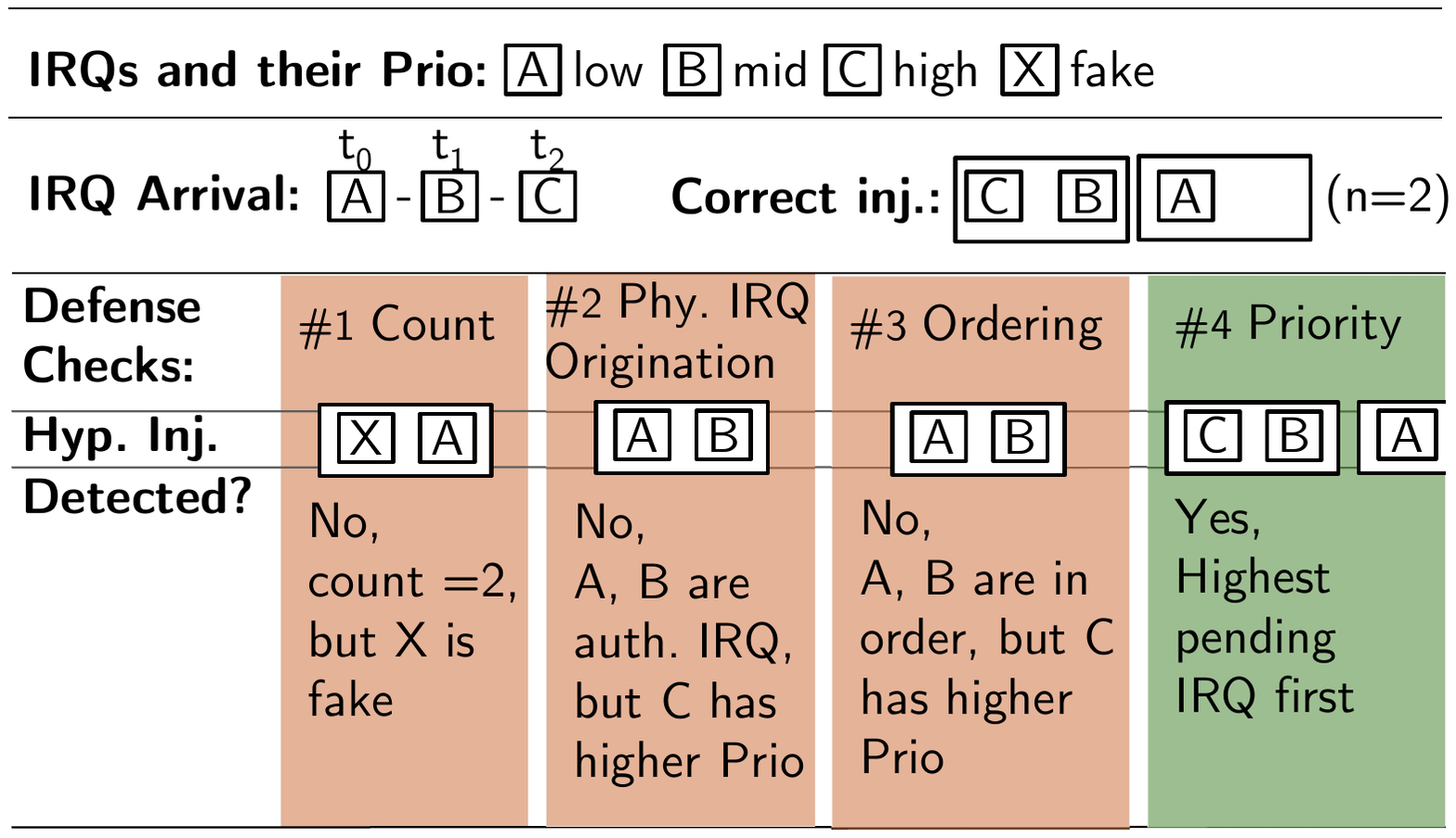}
\caption{
   Device interrupts A,B,C are registered by a \cvm for \codename secure interrupt delivery.
   vGIC window size $n$=2.
   \codename achieves interrupt isolation by enforcing checks \#2, \#3, \#4.
     }
    \label{fig:interrupt-checks}  
\end{figure}

Next, we explain checks that the RMM has to perform
with the \sm cooperation.
We start with simple checks that are insufficient and iterate over  them to build towards our full set of checks.
\cref{fig:interrupt-checks} shows each check with an example.

\newparagraph{Check \#1: Total Count.}
To begin with, assume that the RMM performs a simple count-based check when the hypervisor injects virtual interrupts into a \cvm.
Concretely, the RMM computes the expected number of pending interrupts per \cvm from the \sm recorded physical interrupts and decrements this count on each virtual injection.
This bounds injections to neither too many nor too few (see \cref{fig:interrupt-checks} \#1),
but does not prevent the hypervisor from injecting an interrupt number the device has never raised.

\newparagraph{Check \#2: Origination from a Physical Interrupt.}
To stop the hypervisor from arbitrarily injecting virtual interrupts, the RMM should tie the authenticity of the physical interrupts to the virtual interrupts. 
For this, the RMM requires the \sm to write the registered physical interrupt numbers that arrive to a list in realm memory. 
Then, when the hypervisor injects the virtual interrupt, the RMM ensures that for each virtual interrupt programmed into the vGIC there is a corresponding entry in the \sm list before forwarding it.
In CCA, when the hypervisor injects the virtual interrupt via the RMM, it can send $n$ (where $n$ depends on the GIC implementation) distinct interrupts to the vGIC (see background \cref{sec:background}).
So, for each injection request, the RMM checks that all of the interrupts in the request are authentic. 
Tying the virtual interrupt to physical interrupts stops the hypervisor from arbitrarily injecting other interrupts but cannot ensure interrupt ordering (see \cref{fig:interrupt-checks}~\#2). 

\newparagraph{Check \#3: Ordering.}
\codename needs to store an ordered list of physical interrupts when they arrive at the \sm to guarantee that the hypervisor cannot 
reorder them. Then, the RMM can use this ordered list to check the virtual interrupts that the hypervisor wants to inject. 
We observe that because the hypervisor can inject $n$ distinct interrupts at once to the \cvm, \codename does not need to guarantee ordering for these $n$ distinct interrupts.
If the RMM programs $n$
distinct interrupts into the vGIC, they become pending together on \cvm entry.
So the RMM uses the \sm ordered list to check ordering with a window size of $n$. 
This prevents the hypervisor from reordering the interrupts that are outside this window (\cref{fig:interrupt-checks}~\#3). 
However, this alone cannot achieve \codename interrupt isolation.
During benign operation, the hypervisor injects virtual interrupts based on the priority of all the pending interrupts. 
Thus, time-based ordering cannot preserve the priority. 
\begin{figure}
  \centering
\includegraphics[width=0.9\linewidth, trim=0.5cm 3cm 1cm 3.4cm]{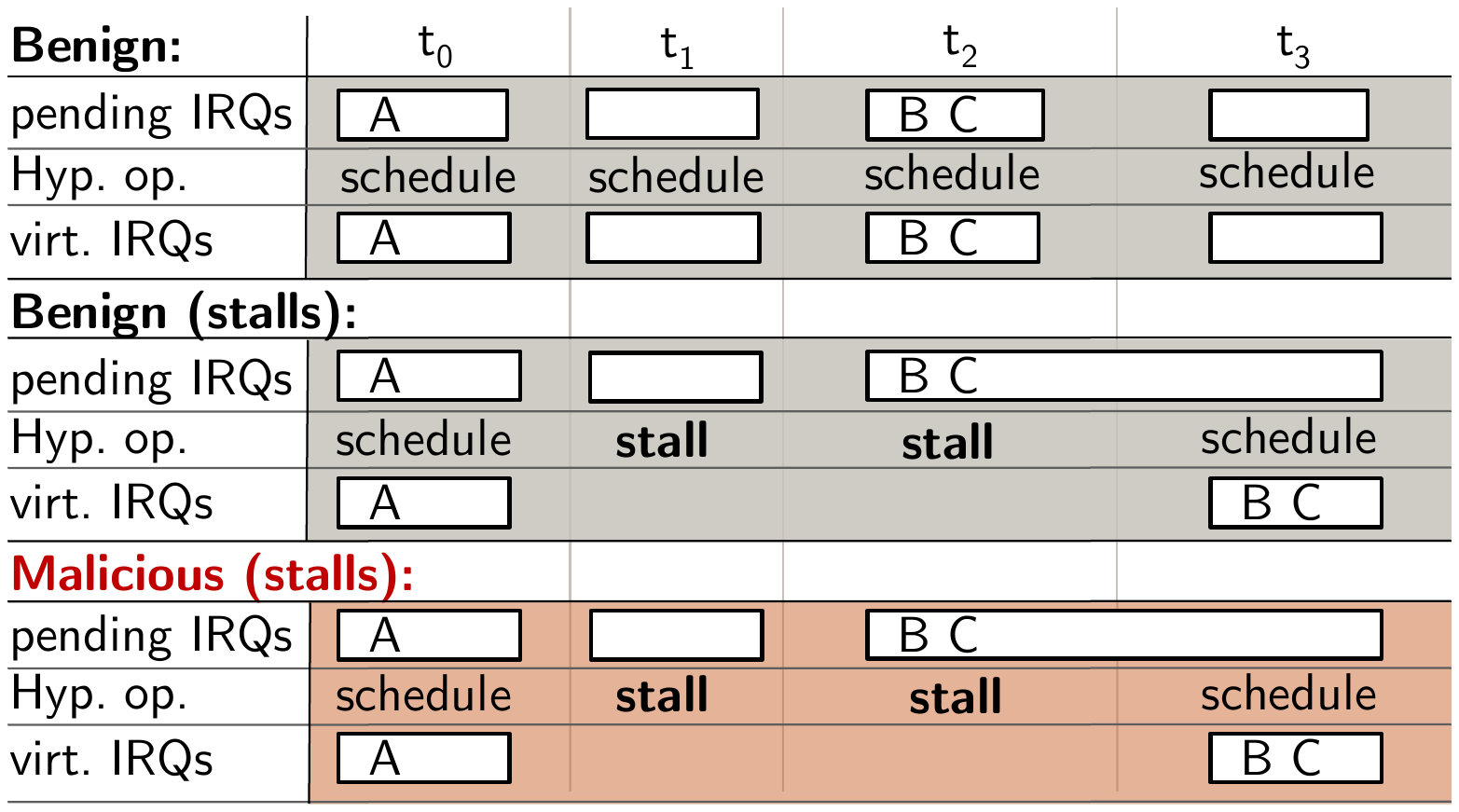}
\caption{
  Top: benign \cvm scheduling.
  Middle: benign scheduling with stalls.
  Bottom: malicious stalls but indistinguishable from middle trace; stalls are also possible under benign scheduling.
  Hyp. op: Hypervisor operation (schedule or stall): A,B,C: interrupts.
  }
\label{fig:malicious-sched}
\end{figure}

\newparagraph{Check \#4: Priorities.}
To check virtual interrupt priorities, \codename requires the \cvm to assign priorities to its registered device interrupts during device attach.
The RMM uses this information to create per-\cvm ordered priority lists and checks the virtual interrupt injections from the hypervisor against them, so that the hypervisor always injects the highest-priority pending interrupts
(\cref{fig:interrupt-checks}~\#4).
This check is sufficient to guarantee that the interrupt behavior observed by a \cvm is equivalent to some benign execution. The hypervisor cannot inject or reorder interrupts in a way the \cvm would not already tolerate.
The guarantee holds even if the hypervisor maliciously
delays \cvm scheduling until a high-priority interrupt arrives to effectively never inject a lower-priority interrupt.
Since a benign hypervisor provides no scheduling guarantees to a \cvm, any malicious delay is indistinguishable from some benign scheduling (\Cref{fig:malicious-sched}).
Together, checks \#2, \#3, and \#4 achieve \codename interrupt isolation.

\subsection{Acknowledging Interrupts}
\label{ssec:ack-int}
The above \codename checks ensure isolated interrupt \config and delivery. 
Despite these checks, the hypervisor still needs to acknowledge interrupts by writing to the memory-mapped registers for functionality. 
\codename marks the memory-mapped register for acknowledgment as root memory, thus revoking the hypervisor access and forcing SMC. 
When the hypervisor tries to acknowledge any interrupt using a \sm call (SMC), the \sm first checks if this interrupt is isolated by \codename. If not, the \sm performs the acknowledgment. 
On the other hand, for interrupts that \codename isolates, it has to ensure that they are acknowledged appropriately for preserving the hypervisor functionality (\cref{appx:design-interrupts}).

Put together, \codename mechanisms for blocking malicious physical injection to establish interrupt authenticity, recording authentic physical interrupts to create the ground truth, and rigorous checking of virtual interrupt injections against the ground truth ensure interrupt isolation for \cvms.

\section{Security Analysis}
\label{sec:security_analysis}

We argue why \codename is secure against different adversaries. 
Prior works provide the guarantees (\ref{g:id}, \ref{g:attach}, \ref{g:mem}, \ref{g:detach}) for device lifecycle management and memory isolation. In \codename we rely on these guarantees for security, as without them \codename interrupt protection will not hold. While we assume that device lifecycle management and memory isolation are secure, for completeness we explain the attacks mitigated by the guarantees in~\cref{appx:sec-analysis}. Next, we explain how the attacks on interrupt isolation are mitigated in \codename.

\newparagraph{GIC Configuration and Interrupt Injection Attacks.}
The hypervisor may try to reprogram the GIC to bypass \sm traps, alter interrupt priorities, or acknowledge interrupts directly, all of which would undermine \codename interrupt security guarantees.
\codename protects the GIC \config in root memory and checks all hypervisor accesses to the \config stopping such attempts. 
The hypervisor can mount Iago attacks on \codename interfaces by invoking the monitor GIC \config interface (as shown in \cref{fig:design:overview}.a) with arbitrary addresses in the root world to read or write from~\cite{checkoway2013iago}. 
\codename stops such attempts by checking that the address range the hypervisor requests to operate on using this interface is in the GIC \config space. 
Besides the memory-mapped GIC \config, the Arm architecture contains system registers for GIC \config. 
The hypervisor can try to use these registers to compromise \codename interrupt isolation. 
However, \codename traps all registered interrupts to EL3.
On Arm, the hypervisor in normal world EL2 cannot use the system registers to configure these EL3 interrupts, thus stopping such attacks. 

The hypervisor can stall \cvm scheduling (\cref{fig:malicious-sched}) and wait for a high priority interrupt to arrive. It can use this malicious scheduling to not inject lower priority interrupts that had arrived earlier. 
However, we observe that this interrupt injection behavior could also occur during \cvm execution with a benign hypervisor as hypervisors do not provide scheduling guarantees for \cvms. 
Thus, this malicious stalling of the \cvm is safe and does not lead to an interrupt injection behavior that diverges from what the \cvm normally expects.
The hypervisor can try to inject an interrupt multiple times (e.g., 16 times) with one vGIC programming request. 
This is not possible, as the vGIC  injects any given interrupt only once on every realm entry.
Therefore, while the vGIC can be programmed to fire $n$ distinct interrupts at once it cannot be programmed to fire the same interrupt multiple times at once. 
Finally, the hypervisor cannot interrupt the RMM mid-check, as the RMM masks all interrupts during execution.

An attacker-controlled co-resident \cvm can try to inject interrupts to other \cvms. 
This is not possible as each \cvm can only program its local vGIC and cannot access vGIC of other \cvms, because the RMM does not expose any APIs for this. 
An attacker can use a device to fake victim device interrupts. 
But the hardware integration of the device on the platform is trusted and is part of attestation report. 
Each device interrupt line is unique and tied to the hardware: software attackers cannot use devices to~fake~other~device~interrupts. 
In~\cref{ssec:sec-eval}, we perform a comprehensive security evaluation on Arm FVP to show that \codename mitigates the attacks discussed here.

\section{Implementation}
\label{sec:implementation}

We prototype \codename on the FVP and change 3054~\loc in the
trusted firmware v2.8 (TF-A) which we use as the \sm~\cite{arm-tfa}, 1188~\loc in RMM
v0.2.0~\cite{arm-rmm}, and 2706~\loc in the Linux kernel v6.2.0 (cca-full/rfc-v1)~\cite{cca-kernel}
which we use as the host and guest OS (guest: 60~\loc, host: 2646~\loc). The Arm reference
implementation uses KVM in the Linux kernel and a virtual machine manager (VMM) called kvmtool to
build and deploy \cvms which we reuse. We change VMM with 590~\loc to add device support.
Our modifications require no application or device-driver changes.
We introduce 6 new interfaces for interrupt protection in \sm and RMM. The full API list is shown in \cref{appx:api-changes}.

\newparagraph{Changes to Platform Boot.}
Like previous works, we create an additional GPT during boot for the device-side GPT (\gptd) to allow devices access to \cvm memory (\cref{appx:impl-attach-detach}). 
To ensure memory isolation, we move the SMMU \config and memory to the root world~\cite{acai,cage}. 
We also move GIC \config to the root world during boot.
Further, we modify the RMM to store platform device information, i.e. interrupt-to-device mappings. During runtime, the RMM uses this information to check \cvm interrupt registrations.

\newparagraph{Interrupt Configuration.}
To isolate physical interrupts, we move the GIC memory-mapped \config space to root. 
So, all hypervisor GIC \configs cause granule protection faults (GPF). 
We modify the GPF handler to invoke the \sm (\smcgicconfig) to check and perform the \config.
We implement an SMC  (\smcprotint) for the RMM to invoke for registering interrupts for isolation with \codename to ensure that they always trap to EL3. 
We first configure the GIC to trap all Group 0 interrupts to the \sm in EL3 and then program the GIC entry for the registered interrupts as belonging to Group 0. 
This ensures that these interrupts always trap to EL3 (\step{1}, \Cref{fig:design:overview}.d).

\newparagraph{Interrupt Delivery.}
We allocate a shared data structure for the RMM 
(realm) and hypervisor (normal) to track pending protected interrupts. 
The \sm populates these for each registered interrupt trap (\steps{2a--b in \cref{fig:design:overview}.d}) and notifies KVM using a software interrupt  (\step{3--4}). 
KVM injects pending interrupts into \cvm (\step{5}) by requesting vGIC system register programming, which the RMM checks and forwards to the \cvm (\step{6}).

\newparagraph{Interrupt Acknowledgment.} 
If possible, (i.e., without causing an interrupt storm), the \sm acknowledges any registered interrupts immediately before invoking KVM (\steps{1-3}).
Otherwise, after the virtual interrupt fires in the guest (\step{7}) and the guest handler completes, the RMM traps the guest and invokes the \sm (\smcackint) to acknowledge the corresponding physical interrupt (\cref{ssec:ack-int} and \cref{appx:design-interrupts}).

\section{Evaluation}
\label{sec:evaluation} 
We evaluate \codename with two complementary prototypes: a functionality prototype on the Arm FVP
and a performance prototype based on OpenCCA~\cite{opencca} and a Rockchip RK3588 Armv8.2 board.
The functionality prototype validates \codename compatibility with Armv9 and Arm CCA, but is limited
to software simulation; the performance prototype estimates \codename overhead on currently
available Armv8 hardware.
As no public CCA hardware is available today, our evaluation, like previous works~\cite{acai,cage,shelter,portal,opencca} provides a best-effort estimate of \codename functionality and performance on future CCA systems.
We defer FVP functionality and compatibility results to \cref{appx:eval} and focus the main text on security evaluation, quantitative performance overheads, and end-to-end case studies.

Across all experiments, we compare a baseline configuration without interrupt protection (\B) against a configuration with interrupt protection and \codename changes (\I).
In both configurations, we boot the platform, launch a realm world \cvm, attach integrated devices, and measure interrupt-related behavior.

\subsection{Security Evaluation on Arm FVP}
\label{ssec:sec-eval}
We validate that \codename enforces correct interrupt isolation with 4~checks.
These checks instantiate adversarial behaviors from \cref{sec:security_analysis} and empirically validate that our implementation enforces \codename security guarantees.

\newparagraph{(a) Benign Interrupts.} We run device workloads (GPU, UART, Mouse \& Keyboard, LED \& Button) that generate up to 10k interrupts
and verify that all these device-raised interrupts reach the \cvm correctly, as expected.

\newparagraph{(b) Malicious Virtual Interrupts.} 
We repeat the benign workloads but additionally modify the hypervisor to inject 10k virtual interrupts that have no corresponding physical interrupt of a protected device.
For this, we create a kernel module that invokes \texttt{kvm\_vgic\_inject\_irq} on the KVM instance of the \cvm (174 \loc).
We report that \codename rejects all injected malicious interrupts in the RMM and only delivers virtual interrupts to the \cvm that are backed by a recorded physical interrupt, as expected.

\newparagraph{(c) Malicious Physical Interrupts.}
We repeat the benign workloads but additionally modify the hypervisor to write to the GIC configuration to raise 10k physical interrupts for a protected device (Set-Pending register, \texttt{GICD\_ISPENDR} on the Distributor MMIO of the GIC, 168 LoC).
We report that these malicious writes trap to the monitor and are blocked from reaching the GIC, as expected.

\newparagraph{(d) Malicious Memory Accesses.}
We configure the FVP SMMU test engine~\cite{fvp-testengine} to issue memory accesses to the MMIO region of the protected device, including the registers for interrupt acknowledgement.
The test engine is not attached to the \cvm, so all accesses are unauthorized. We report that our implementation blocks these accesses with GPT violations.
In addition, we modify the hypervisor to read from the MMIO region of the protected device directly, and report that these accesses are likewise rejected with GPT faults (176 \loc).

\begin{figure*}[t]
\centering
    \includegraphics[width=0.99\textwidth, trim=2.9mm 7mm 6mm 2mm, clip]{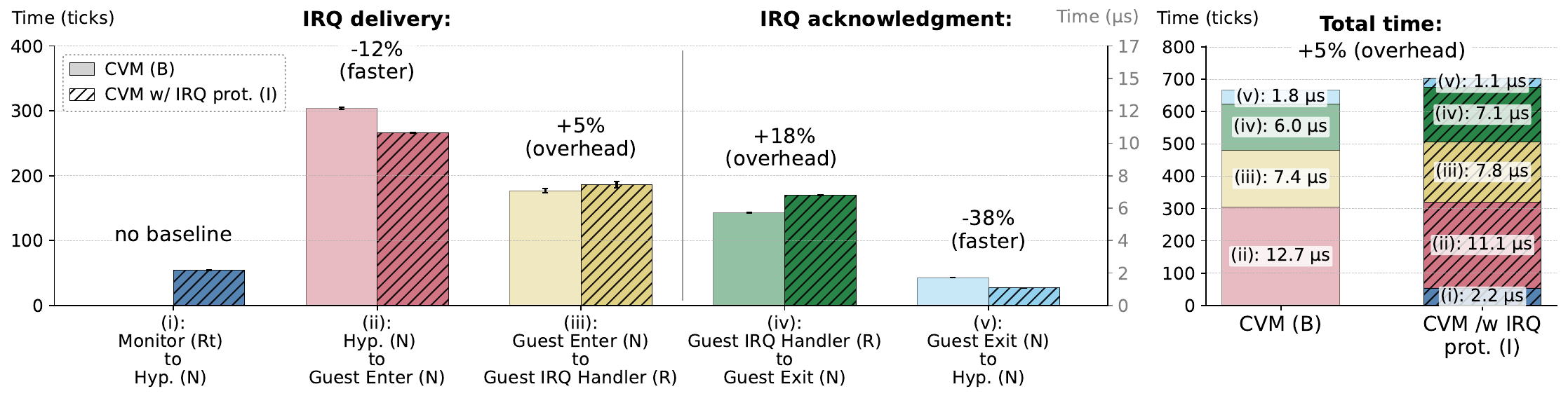}     
         \caption{
GPU interrupt-flow breakdown on the Arm board under \codename.
Left: stage times for
(i) \sm IRQ trap and hypervisor notification (Steps 1-4 in design \Cref{fig:design:overview}),
(ii) hypervisor schedules guest (Step 5 in \Cref{fig:design:overview}),
(iii) guest entry, RMM checks, IRQ handler (Steps 6-7 in \Cref{fig:design:overview}),
(iv) guest exits back to hypervisor,
(v) end-of-interrupt acknowledgement in hypervisor.
Bars show baseline CVM \B and CVM with \codename IRQ protection (\I);
percentages indicate overhead of \I over \B.
No baseline exists for (i), since without \codename the IRQ directly reaches (ii).
Right: total interrupt-handling time across (i)-(v).
N: Normal, Rt: Root, R: Realm. 1 tick = 41.67 ns. Lower is better.
}
    \label{fig:gpu:latency}
\end{figure*}

\subsection{Performance Prototype}
We estimate \codename overheads on an Armv8 board with OpenCCA~\cite{opencca}.
We apply \codename on top of OpenCCA tree (TF-A v2.11, RMM v0.5, Linux v6.12)~\cite{opencca-version}
and run on a Rock 5B RK3588 SoC, 16~GB of RAM, and Mali G610.
We boot the platform on 4 Cortex-A55 cores at 1.8~GHz (\texttt{userspace} governor).

\newparagraph{Impact on Boot and Normal Operations.}
\codename adds 2.4\% overhead (221\,ms) to the platform boot.
This increases total boot time from 8.9\,s to 9.1\,s.
Most of this overhead comes from SMMU setup for memory isolation (+177.7\,ms), while \codename's GIC configuration (i.e., protecting the GIC in root world) adds only 0.1\,ms during BL31.
During Linux boot with \I, we observe 1816 traps (445 reads, 1371 writes) for protected GIC configuration; each trap takes 2.5\,\textmu s, which is negligible relative to the 4.8\,s kernel boot time.
To evaluate non-interrupt workloads, we run sysbench~\cite{sysbench} on a normal world VM and a \cvm under \B and \I.
As expected, \codename has no measurable impact on workloads that do not involve device interrupts (details in \Cref{appx:sysbench}).

\newparagraph{Interrupt Lifecycle Breakdown.}
We measure the cost of a single protected interrupt, from when the device raises a physical interrupt until the hypervisor processes the end-of-interrupt acknowledgement,
and break this path into interrupt delivery (\circnum{1}--\circnum{3}) and acknowledgement (\circnum{4}--\circnum{5}) in \Cref{fig:gpu:latency}.
We use the Mali G610 GPU on the Arm board, 
which generates frequent interrupts and requires a complex driver stack~\cite{nomali-testsuite}.
Averaged across 1000 interrupts, the end-to-end interrupt time in \B is 669 ticks (27.9\,\us, 95\% CI: $\pm$0.2\,\us), while \I increases this to 703 ticks (29.3\,\us, 95\% CI: $\pm$0.2\,\us). This corresponds to a 5\% overhead.
The stage-wise breakdown in \Cref{fig:gpu:latency} shows that this
overhead comes from recording in the \sm \circnum{1}, during
authenticity checks in the RMM on guest entry \circnum{3}, 
and guest exit for end-of-interrupt acknowledgement in the \sm \circnum{4}.
This is expected because \codename adds bookkeeping and checks in these stages.
In contrast, \codename is faster when the hypervisor schedules the guest with the new virtual interrupt \circnum{2},
  and when it completes post end-of-interrupt bookkeeping \circnum{5}.
This is because, with \codename, the \sm manages the physical interrupt state (disable on trap, re-enable on acknowledgment) rather than leaving it to the~hypervisor.

\begin{figure*}[t]
\centering
    \includegraphics[width=0.99\textwidth, trim=2.8mm 1mm 3.2mm 1mm, clip]{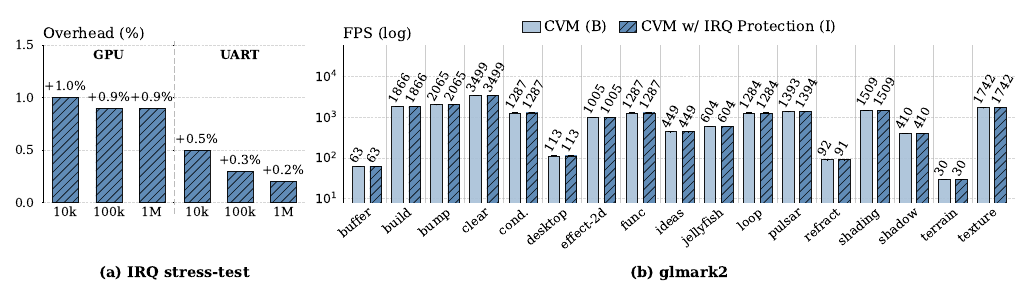} 
         \caption{
         Arm-board evaluation. Left: GPU and UART IRQ stress test; Y-axis shows overhead of \I over \B. Lower is better. Right: glmark2 Frames/sec (GPU) for \B and \I in log scale. Higher is better.
         }
    \label{fig:gpu:glmark2}
\end{figure*}

\newparagraph{Sustained Interrupt Load.}
We examine whether these overheads accumulate under sustained device activity.
For this, we run an interrupt stress test using both the GPU and UART device.
We attach each device to their own \cvm and configure the device to generate a high frequency of interrupts.
For GPU, we use Mali KUTF Midgard benchmark suite~\cite{nomali-testsuite}.
For UART, we modify the Synopsys DesignWare 8250 driver to run in internal loopback mode. This way,
it raises a physical interrupt on each write to the transmit register.
Under continuous load, the overhead stays within 1\% (stdev below 1.2\% of mean). This indicates that per-interrupt costs do not accumulate at scale (see \Cref{fig:gpu:glmark2}, left).

\subsection{Case Studies}
We validate \codename with real-world apps on FVP and board.

\newparagraph{FVP.} %
On FVP, we run three interactive UART-based applications inside a \cvm: a terminal browser, an IRC client and a text editor. 
Because interactive actions are difficult to reproduce consistently across runs, we use these experiments only as qualitative validation. This demonstrates \codename compatibility and usability compared to executing them in the realm world without interrupt protection.
We provide videos for both \B and \I in~\cite{video-casestudy-br,video-casestudy-dmi}; details in \Cref{appx:case-studies}.

\newparagraph{Board.} 
We evaluate our design under realistic GPU workloads using glmark2 \cite{glmark2}, a suite of userspace workloads that capture the mixed compute, memory, and driver interactions of typical integrated GPU apps. The workloads generate 2.5k~--~32k GPU interrupts (configuration in \Cref{appx:casestudy:board:details}).
Enabling \codename interrupt protection across all 17 benchmarks has no measurable impact on the end-to-end throughput (see \Cref{fig:gpu:glmark2}, right). Frame rates for \cvm with and without \codename are almost identical.
The geometric mean slowdown across all benchmarks is 0.06\%. 
So, \codename interrupt protection does not add measurable overhead beyond the baseline realm cost, indicating no end-to-end performance overheads for \cvms.

\section{Related Work}
\label{sec:related_work}

\newparagraph{Arm CCA and Device Protection.}
Prior works on Arm CCA~\cite{cage,acai,portal}
\cite{castes2024sharing,shelter,via,tfx,zhang2025scrutinizer,ye2024fortifypatch,zhou2025rcontainer,chen2024cubevisor,chen2024cpc},  
offer better abstractions~\cite{zhou2025rcontainer,chen2024cpc,chen2024cubevisor,castes2024sharing} or new security guarantees \cite{zhang2025scrutinizer,ye2024fortifypatch}.
Cage and Strongbox build support for CCA \cvms and TrustZone trusted apps respectively, to securely
compute on integrated Arm GPUs---without trusting the GPU driver and runtime to reduce the
TCB~\cite{cage,strongbox}.
\codename instead keeps drivers and runtimes in the \cvm TCB and, like Portal, supports a wide array of integrated devices in \cvms~\cite{portal}.
Similarly, Acai connects TEE-enabled PCIe accelerators to \cvms~\cite{acai},
and Arm hardware extension for Device Assignment (RME-DA) allows such accelerators
to directly access \cvm memory~\cite{rme-da}. 
Neither applies to integrated devices, but \codename reuses techniques from Portal, Acai, and RME-DA for
device lifecycle and memory isolation, and extends them with interrupt isolation as its new contribution.
TrustZone-based works enable protected device access~\cite{teetime,trustice,trustotp,VButton,TruZ-Droid,trustUI,SeCloak,han2023mytee},
but do not reason about VMs or virtual interrupt delivery.
\codename targets CCA and reasons about multiplexing interrupts across CVMs.
Ongoing verification of CCA firmware~\cite{via,tfx} can be extended to \codename.

\newparagraph{Integrated Device Protection in Other Architectures.}
On RISC-V, prior works attach devices to enclaves with hardware changes~\cite{cure,schneider2022composite}.
\codename uses CCA's analogous hardware-based isolation.
Others let trusted hardware create different views of memory like GPTs in CCA~\cite{IOPMP,siopmp,brasser2019sanctuary} or hardware-domains~\cite{yao2023minimizing}.
Intel SGX-based trusted I/O~\cite{weiser2017sgxio,aion,fidelius,sgx-usb,proximitee}
and prior enclave defenses against malicious non-device interrupts~\cite{de2014secure,busi2021securing} target userspace enclaves,
whereas \codename isolates device interrupts for \cvms.

\newparagraph{Interrupt Attacks.}
Side-channel attacks exploit unprotected timers, page-faults, IPIs~\cite{nemesis,frontalattack,van2017telling,xu2015controlled,he2018sgxlinger,sgxfail,sgx-pf-sidechannel,crosscoreipi}, and fake CPU exceptions, system calls, and exploit race conditions~\cite{heckler-usenix,wesee-oakland,ExpRace} 
These interrupts are raised by the CPU (e.g., page-faults, system calls, IPIs), whereas \codename focuses on device interrupts. 
Prior works have studied how statistical information on integrated device interrupts (e.g., interrupt timing, \# interrupts) can compromise~confidentiality~\cite{no-pardon}.

\newparagraph{Interrupt Isolation on Arm.}
For {\em interrupt configuration}, prior works on TrustZone~\cite{teetime,strongbox,SeCloak} isolate integrated device interrupts by marking the GIC \config as secure world.
However, in \codename, we do not trust the secure world which can directly program the GIC to maliciously fire interrupts. So, departing from these works and for security in the CCA setting, \codename marks GIC \config as root.
On the other hand, for {\em interrupt delivery}, like the above  prior works, \codename also marks interrupts as belonging to Group 0 (i.e., secure world) to trap them in EL3~\cite{strongbox,teetime,SeCloak}.
Recent CCA works need interrupt isolation for one interrupt ID (e.g., GPU or timer) and resort to the same mechanisms as works on TrustZone~\cite{cage,hitchhiker} i.e., marking the interrupt as secure world Group. 
Thus, they have to trust the secure world, which \codename avoids. 
Other works on Arm CCA have proposed delegating interrupt management to the RMM for performance when defending against side-channel adversaries~\cite{castes2024sharing}. 
In their design, they pin \cvms to cores and reduce the number of context switches to the hypervisor by using the RMM to manage the \cvm timer interrupts. 
This is equivalent in security to the existing CCA threat model that does not block  malicious interrupt injection from the hypervisor. 
As the hypervisor still manages all other interrupts, their design would need \codename interrupt isolation. 
\codename is not specific to one interrupt ID, instead isolates device interrupts across multiple devices and CVMs thus addressing the gap in the CCA threat model. 

\newparagraph{TDISP, Intel TDX, and AMD SEV-SNP.}
TDISP is a PCIe feature that allows direct attachment of TEE-enabled devices to CVMs and does not reason about device interrupts.
TDX-Connect and SEV-TIO are TDISP-compliant implementations specific to Intel and AMD, respectively.
They protect CVM-bound interrupts by deploying trusted firmware and software to check virtual interrupt injections~\cite{tdx,sev-snp,svsm-github}. 
But, like CCA, they only allow or deny all device interrupts without granular interrupt isolation. 
\codename insights can be applied to these other architectures, but with careful analysis.

\section{Conclusion}
In this paper, we present \codename, a device interrupt isolation mechanism for \cvms. 
Our delegate-but-check strategy for interrupt isolation allows the hypervisor to manage the GIC but strictly monitors the interrupt lifecycle (\config, delivery, acknowledgment).
Our evaluation demonstrates this on the Arm FVP and an Arm Rock 5B Board.

\bibliographystyle{splncs04}
\bibliography{paper}

\appendix
\crefalias{section}{appendix}
\crefalias{subsection}{appendix}
\crefalias{subsubsection}{appendix}

\section{Security Analysis: Memory Isolation}
\label{appx:sec-analysis}
We detail attacks on device lifecycle management and memory isolation mitigated by guarantees (\ref{g:id}, \ref{g:attach}, \ref{g:mem}, \ref{g:detach}) \codename assumes from prior works.

\newparagraph{Attach/Detach Device.} 
The untrusted hypervisor can try to map a device to a physical address space it controls before attaching it to a \cvm. 
This is not possible as the physical addresses that the devices are mapped to are fixed for the hardware platform and cannot be changed by software (\ref{g:id}).
Next, a hypervisor can try to emulate devices or assign the wrong device to a \cvm which is stopped by ensuring unforgeable device identities (\ref{g:id}). 
Concretely, \codename checks the physical addresses that the hypervisor maps for the device using trusted platform information in the \sm to detect and stop such attacks.
The hypervisor can also try to assign a device in a state that compromises the \cvm when it is attached to it. 
To stop such attacks, \codename soft resets the device before attaching it to the \cvm (\ref{g:attach}).
Similarly, the hypervisor can request \codename to attach a device to a \cvm when the \cvm does not expect to have a device; this is stopped by \ref{g:attach}. 
\codename does not allow such requests and expects the \cvm to request a device attach.
Finally, the hypervisor can try to detach a device while a \cvm is using it to leak device data; this is mitigated by~\ref{g:detach}.

\newparagraph{Co-Resident \cvms.}
Attacker-controlled co-resident \cvms can try to directly access the device MMIO or DMA memory. 
An attacker can also try to set up realm memory regions overlapping with device memory of the victim \cvm, but this would violate~\ref{g:mem}. 
\codename adds all device memory to the \cvm. 
When memory is added to a \cvm, CCA ensures that this memory only belongs to one \cvm and isolates it by programming the \stables in the RMM. 
Thus, other \cvms will not have valid mappings for the device memory and cannot access it.

\newparagraph{SMMU Configuration and Attacker-controlled Devices.}
Hypervisor can try to change SMMU \stable mappings to allow attacker-controlled devices access to victim device memory but this would violate~\ref{g:mem}. 
To stop these attacks and guarantee~\ref{g:mem}, \codename protects the SMMU \config in the root world to check and stop these attempts. 
Malicious devices can try to directly access victim device memory, but are stopped by \ref{g:mem}.
Concretely, \codename ensures that the attacker-controlled device \stables will not have valid mappings to access victim device memory.

\section{Implementation Details}

\begin{table*}[t]
    \centering
    \caption{\codename new and existing APIs for device lifecycle management,
    memory isolation, and interrupt isolation. RSI and RMI are implemented in
    the RMM. SMCs are implemented in \sm.}
    \label{tab:api-changes}
\small
    \begin{tabular*}{\textwidth}{
        @{\extracolsep{\fill}}
        lll
        @{}
    }
        \toprule
        \textbf{API} & \textbf{Status} & \textbf{Description} \\
        \midrule
        \rsiattachdev
            & New
            & Request attach device to \cvm \\

        \rsidetachdev
            & New
            & Request detach device from \cvm \\

        \rmidevfinalise
            & New
            & Hyp. finished adding device memory to realm \\

        \rmidatacreate
            & Modify
            & Update SMMU if devices attached to \cvm \\

        \rmidatadestroy
            & Modify
            & Update SMMU if devices detached from \cvm \\

        \rmigranuledelegate
            & Modify
            & Update 2 GPTs \\

        \rmigranuleundelegate
            & Modify
            & Update 2 GPTs \\

        \smcackint
            & New
            & Acknowledge interrupt \\

        \smcgicconfig
            & New
            & R/W to GIC \config \\

        \smcprotint
            & New
            & Mark interrupts as protected \\
        \bottomrule
    \end{tabular*}
\end{table*}

\subsection{Device Attach and Detach}
\label{appx:impl-attach-detach}
\label{ssec:attach-detach-device}
\begin{figure}[t]
  \centering
\includegraphics[width=1\linewidth, trim=0cm 0cm 0cm 0cm]{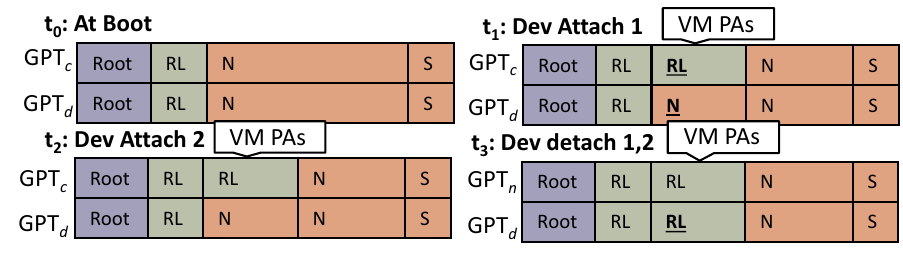}
\caption{\codename's 2 GPTs. t0: Create and populate 2 GPTs (\gptc for cores, and \gptd for devices). t1: Update \gptd on device attach to allow devices access to \cvm memory by marking \cvm as normal world (N) in \gptd. t2: No updates. t3: Update \gptd to revoke device access to \cvm memory by marking it back to realm world (RL).}
\label{appx:fig:gpts}
\end{figure}

\codename adapts the device attach process of prior works to allow \cvms to enable \codename interrupt isolation~\cite{portal,rme-da,acai}.
CCA does not allow devices direct access to \cvm memory.
As in those works, \codename uses 2 GPTs to attach devices to \cvms (\cref{appx:fig:gpts}). 
We create \gptc for the cores and 
\gptd for the devices (i.e., the SMMU GPT).
In \gptd, \cvm memory shared with the device is marked as normal world, while in \gptc the same memory is marked as realm world.
\label{appx:api-changes}

\newparagraph{Attach.}
To allow \cvms to register specific device interrupts we implement a new RSI call (\rsiattachdev) in the guest Linux kernel and the RMM, which the \cvm invokes with the device details (i.e., ID, memory-mapped GPAs, registered device interrupts). 
The RMM checks this request by looking up platform information (e.g., physical address ranges for device MMIO, device interrupt numbers) and then forwards the request to the hypervisor. 
Then, the hypervisor transitions all memory for the device to realm world (\rmigranuledelegate), adds this memory to the \cvm (\rmidatacreate), and invokes \rmidevfinalise. 
In response, the RMM checks that the hypervisor has installed the correct mappings for the \cvm based on the \cvm request. 
For memory isolation, the RMM invokes the \sm to configure the SMMU and marks the \cvm memory as normal world in \gptd and soft resets the device. 
Then, for registered interrupts, the RMM invokes the \sm (\smcprotint) to configure it and returns to the \cvm indicating that the device attach is complete. 

\newparagraph{Detach.}
The \cvm can request a device detach (\rsidetachdev). \Cref{tab:api-changes} summarizes the APIs.

\begin{table*}[t]
\centering
\caption{Device-driver compatibility on the Arm FVP. \codename supports
unmodified drivers in \cvm for all tested peripherals. IRQ indicates whether
the driver uses interrupts. The LED and button do not generate interrupts and
therefore do not require interrupt protection.}
\label{tab:device-compat}

\small
\begin{tabular*}{\textwidth}{
    @{\extracolsep{\fill}}
    l
    c
    c
    c
    l
    @{}
}
\toprule
\textbf{Device} &
\makecell{\textbf{IRQs}} &
\makecell{\textbf{Driver}\\\textbf{Mod.}} &
\makecell{\textbf{Protected}\\\textbf{w/ \codename}} &
\textbf{Setup \& Benchmark} \\
\midrule
GPU
    & \cmark
    & No
    & \cmark
    & G76 kbase driver, Midgard test suite \\

UART
    & \cmark
    & No
    & \cmark
    & Amba-pl011 driver (nano, browser, IRC) \\

PS2/Mouse \& Keyboard
    & \cmark
    & No
    & \cmark
    & Amba-kmi driver (move mouse, type keys) \\

LED \& Button
    & \xmark
    & No
    & N/A
    & Read button state, flash LED \\
\bottomrule
\end{tabular*}
\end{table*}

\subsection{Interrupt Acknowledgment}
\label{appx:design-interrupts}
A benign hypervisor typically acknowledges physical interrupts as soon as it receives an interrupt and then injects a virtual interrupt to the VM  i.e., before servicing the interrupt. This is to ensure progress in the system---if the acknowledgment is delayed for a particular interrupt, all further interrupt invocations for this id will be blocked. 
However, the benign hypervisor should not use the above greedy acknowledgment approach for special types of interrupts. Specifically, certain devices constantly re-trigger the same interrupt if an interrupt is acknowledged too early i.e., the handler in the VM has not performed its operations. 
For example, the keyboard writes pending characters to its own memory and sends an interrupt. However, if the physical interrupt is acknowledged before the VM handler consumes the characters the keyboard thinks that the interrupt was dropped and resends it. 
If the gap between the hypervisor's greedy acknowledgment and the VM's end of interrupt handler is large, the interrupts keep arriving at the GIC, which injects them into the unblocked cores one after the other. 
This is referred to as an interrupt storm~\cite{interrupt-storms}.

The hypervisor knows the type of interrupt from the platform device information and uses this information to decide if it is safe to perform greedy acknowledgments for cases where it is safe. If it is not safe, the hypervisor first services the interrupt and injects a virtual interrupt to the VM.
Then, the hypervisor waits for the virtual interrupt acknowledgment from the VM and only then performs a physical interrupt acknowledgment. 
So, when the hypervisor requests to acknowledge registered interrupts, \codename relies on the device tree information. 
Based on the platform device information, \codename either allows greedy acknowledgment  or waits for the \cvm to acknowledge the virtual interrupt and only then allows a physical interrupt acknowledgment (\smcackint).

\section{Evaluation Details}
\label{appx:eval}
To evaluate the functionality and security of \codename, we prototype it on Arm's simulator i.e., Fixed Virtual Platform (FVP). In this appendix we detail the setup and results of our evaluation. Then, we provide additional details of our performance evaluation on the Arm board.

\subsection{FVP Setup}
\label{appx:eval:fvp}
We validate \codename 
on the Arm Fixed Virtual Platform (FVP) version \texttt{Base RevC 2xAEMvA 11.20\_15 Linux64}.
We configure it with 4 AEM cores with RME and 3 GB of RAM.
The FVP is a simulator without accurate microarchitectural behavior, so we only validate \codename functionality and compatibility~with~it, not performance.

\subsection{Device Setups on FVP}
\label{appx:eval-devices:fvp}
We attach the following devices to the FVP to test \codename compatibility with unmodified device drivers. The next paragraphs
describe the devices and their workloads (\Cref{tab:device-compat}).
Because the FVP is not cycle accurate, and because asynchronous interrupts like timers can occur irregularly, neither cycle nor event counts are a reliable performance metric. We use FVP's Model Trace Interface (MTI) to count events.
As per our experimental measurements on the FVP, in the baseline \B, guest entry and exit follow a 1:1 ratio: each RMI entry has a matching SMC exit invocation. With \I, this changes because \codename interrupt handling adds extra \sm calls: for each interrupt, the RMM invokes the root-world monitor to acknowledge the interrupt.

\newparagraph{Keyboard.} We use the FVP X11 visualization component that translates keyboard and mouse input on the x86 host to an emulated PL050 PS/2 peripheral on the FVP. We eliminate typing speed effects by using an x86 helper application that uses the X11/Xlib library to control keyboard input. For each key press, the FVP generates a physical interrupt. We run workloads that generate 1000 and 10000 interrupts. 

\newparagraph{GPU.} The FVP supports a limited non-functional implementation of a Mali G76 GPU (No-Mali)~\cite{nomali}. The GPU cannot perform any computation but only exposes an MMIO layout and can send interrupts. So, we run the official Midgard Mali KUTF IRQ test suite~\cite{nomali-testsuite}. The test suite triggers an MMIO write via the Mali driver and prompts the GPU to send 10000 interrupts. 

\newparagraph{UART.} We attach the FVP UART PrimeCell PL011 peripheral to the VMs in our setup. This device facilitates serial communication which we control using an X11/Xlib library. We trigger workloads that generate 1000 and 10000 interrupts. 

\newparagraph{LED \& Button.}
The FVP includes 8 LEDs and 8 User DIP switches (buttons) which can be controlled by software using system registers \texttt{SYS\_LED} and \texttt{SYS\_SW} respectively. 
To test their functionality, we write 1000 random values to these registers and measure the total number of interrupts. 

\newparagraph{Mouse.}
On the FVP, the mouse works by capturing the mouse input in the FVP visualization component. Moving the mouse from the left to the right side of the screen raises 53 interrupts. 
A single mouse click triggers $2$ interrupts. 
While we can verify that the mouse is functional, we cannot benchmark it as we are unable to perform this operation (for moving the mouse, or clicking) without user interaction.

\begin{table}[t]
\caption{Glmark2 with all scenes in their standard configuration for 20k frames.}
\label{tab:glmark2-scenes}
\centering
\resizebox{1\columnwidth}{!}{%
\begin{tabular}{llr}
\toprule
\textbf{Scene} & \textbf{Description} & \makecell{\textbf{\# IRQs}} \\
\midrule

buffer        & Vertex buffer object throughput                 & 2528  \\
build        &  Shader compilation  and setup            & 18678  \\
bump          & Per fragment lighting with bump mapping         & 20030 \\
clear         & Framebuffer clear performance                   & 20035 \\
conditionals  & Shader branching and flow control               & 12906 \\
desktop       & Window compositing and blending effects         & 24793 \\
effect2d      & 2D post processing filters                      & 10085 \\
function      & Function call overhead in shaders               & 12902 \\
ideas         & Dynamic multi object scene rendering            & 14977 \\
jellyfish     & Animated mesh deformation workload              & 6164  \\
loop          & Shader loop execution stress test               & 12886 \\
pulsar        & Procedural animation and texture sampling       & 14212 \\
refract       & Environment mapped refraction rendering         & 25394 \\
shading       & Lighting and material computation cost          & 15124 \\
shadow        & Shadow mapping performance                      & 16351 \\
terrain       & Height mapped terrain rendering                 & 32335 \\
texture       & Texture sampling bandwidth                      & 17514 \\
\bottomrule
\end{tabular}
}
\end{table}

\subsection{Case Study Setup on FVP}
\label{appx:case-studies}

Below we detail our setup for each of the FVP case studies.

\newparagraph{Terminal Based Browser.} We cross-compile Lynx browser v2.9.2 \cite{lynx} for Arm. We attach the UART to the realm VM and browse en.wikipedia.org in Lynx. On the landing page, we navigate to the \textit{From Today's featured Article} section and follow 4 subsequent links~\cite{wiki-today-article}.

\newparagraph{IRC Client.} We cross-compile a Console IRC Client \cite{irc-client} and connect to Miniircd Server v2.3~\cite{miniircd}. The server has 2 other users logged in. We exchange 6 messages with a user.

\newparagraph{Text Editor.} We cross-compile Nano text editor v8.1 \cite{gnu-nano} and add a new DNS entry to \texttt{/etc/hosts}.

\subsection{Case Study Setup on Board}

\label{appx:casestudy:board:details}
We cross-compile glmark2~\cite{glmark2} version 2023.01 for aarch64 Linux and run it in surfaceless mode on the Mali G610 GPU (OpenGL ES 3.1, Mesa 25.0.7-2).
All 17 scenes execute unmodified in their standard configuration at 1920x1080 resolution for 20'000 frames and 30 runs in each environment:
a realm-world \cvm (\B) and a realm-world \cvm with \codename interrupt protection (\I). The GPU frequency is fixed to 800 MHz, and DVFS is disabled to avoid frequency-scaling effects.
We use glmark2 in the default configuration with: buf=32, r=8, g=8, b=8, a=8, depth=24, stencil=0, multisample anti-aliasing=0. \Cref{tab:glmark2-scenes} enumerates the benchmarks.

\subsection{Non Interrupt benchmark Setup}
\label{appx:sysbench}
\begin{figure}
  \centering
        \centering
     \includegraphics[width=1\columnwidth,trim=0 0 0 0cm ]{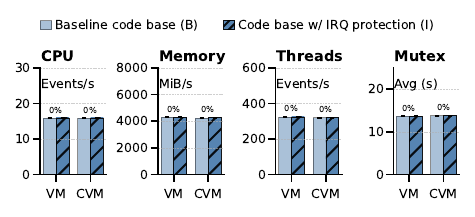} 
         \caption{Sysbench Benchmark on Arm board. No measurable impact on non interrupt workloads.}
    \label{fig:sysbench}
\end{figure}

We run sysbench v1.0.20~\cite{sysbench} on Arm board (CPU: 200k Primes, Memory: 1~GB (64k Blocks), Threads: 8, Mutex: 8) in both a \cvm and normal world VM with (\I) and without (\B) \codename changes.
As expected, since no interrupts are involved, \codename changes incur no measurable overheads (see \Cref{fig:sysbench}).

\end{document}